\documentclass[pra,twocolumn,showpacs,preprintnumbers,amssymb,tightenlines,epsfig,superscriptaddress]{revtex4}
\usepackage{graphicx}
\usepackage{color}
\usepackage[percent]{overpic}

\usepackage{amsmath}
\newcommand{\bra}[1]{\langle\,{#1}\, |}
\newcommand{\ket}[1]{|\,{#1}\,\rangle}
\newcommand{\braket}[2]{\mbox{$\langle\,{#1}\, | \,{#2}\,\rangle$}}

\newcommand{\rvec}[1]{{\mathbf{r}}}

\newcommand{\sub}[2]{{#1}_{\mbox{\!\! \scriptsize #2}}}

\def\beq{\begin{equation}}
\def\eeq{\end{equation}}

\def\CR{\nonumber\\[0.15cm]}

\newcommand{\fref}[1]{Fig.~\ref{#1}}
\newcommand{\frefp}[2]{Fig.~\ref{#1}~(#2)}

\newcommand{\eref}[1]{Eq.~(\ref{#1})}

\newcommand{\sref}[1]{section~\ref{#1}}

\newcommand{\cref}[1]{chapter~\ref{#1}}

\newcommand{\Cref}[1]{Chapter~\ref{#1}}

\newcommand{\aref}[1]{appendix~\ref{#1}}
\newcommand{\bref}[1]{(\ref{#1})}

\usepackage{ulem}  
\begin{document}
%
%
\title{Non-adiabatic dynamics in Rydberg gases with random atom positions}
\author{Ritesh Pant}
\affiliation{Department of Physics, Indian Institute of Science Education and Research, Bhopal, Madhya Pradesh 462 066, India}
\author{Rajat Agrawal}
\affiliation{Max Planck Institute for the Physics of Complex Systems, N\"othnitzer Strasse 38, 01187 Dresden, Germany}
\author{Sebastian~W\"uster}
\affiliation{Department of Physics, Indian Institute of Science Education and Research, Bhopal, Madhya Pradesh 462 066, India}
\author{Jan-Michael Rost}
\affiliation{Max Planck Institute for the Physics of Complex Systems, N\"othnitzer Strasse 38, 01187 Dresden, Germany}
\begin{abstract}
Assemblies of highly excited Rydberg atoms in an ultracold gas can be set into motion by a combination of van-der-Waals and resonant dipole-dipole interactions. Thereby, the collective electronic Rydberg state might change due to non-adiabatic transitions, in particular if the configuration encounters a conical interaction. For the experimentally most accessible scenario, in which the Rydberg atoms are initially randomly excited in a three-dimensional bulk gas under blockade conditions, we numerically show that non-adiabatic transitions can be common when starting from the most energetic repulsive BO-surface. We outline how this state can be selectively excited using a microwave resonance, and demonstrate a regime where almost all collisional ionization of Rydberg atoms can be traced back to a prior non-adiabatic transition. Since Rydberg ionisation is relatively straightforward to detect, the excitation and measurement scheme considered here renders non-adiabatic effects in Rydberg motion easier to demonstrate experimentally than in scenarios considered previously.
\end{abstract}

\maketitle

\section{Introduction} 
%
Collections of a small number of highly excited Rydberg atoms in a cold gas or even BEC are nowadays routinely created in laboratories. The motion of these atoms can often be neglected on the time scale of experiments, justifying the so-called frozen gas approximation. However, under generic circumstances, namely for lighter atoms and initial close proximity, they are easily set into motion on time-scales of interest, by non-resonant van-der Waals (vdW) or resonant dipole-dipole (DD) interactions \cite{gensemer1998ultracold,li2005dipole, fioretti1999long, overstreet2007photoinitiated,noordam:interactions}. The motion of a randomly distributed assembly of Rydberg atoms in three dimensions (3D) has been studied in experiments, however, so far mainly due to interactions involving a single Rydberg state at a time \cite{viteau2008melting,amthor2007mechanical,amthor:vanderwaals,celistrino_teixeira:microwavespec_motion, thaicharoen2015measurement}, permanent dipoles  \cite{Goncalves_permdipole_motion,thaicharoen:dipolar_imaging} or transition dipole-dipole interactions of two Rydberg atoms \cite{park2011dipole,park2011ionization}.

\begin{figure}[htb]
\includegraphics[width=0.99\columnwidth]{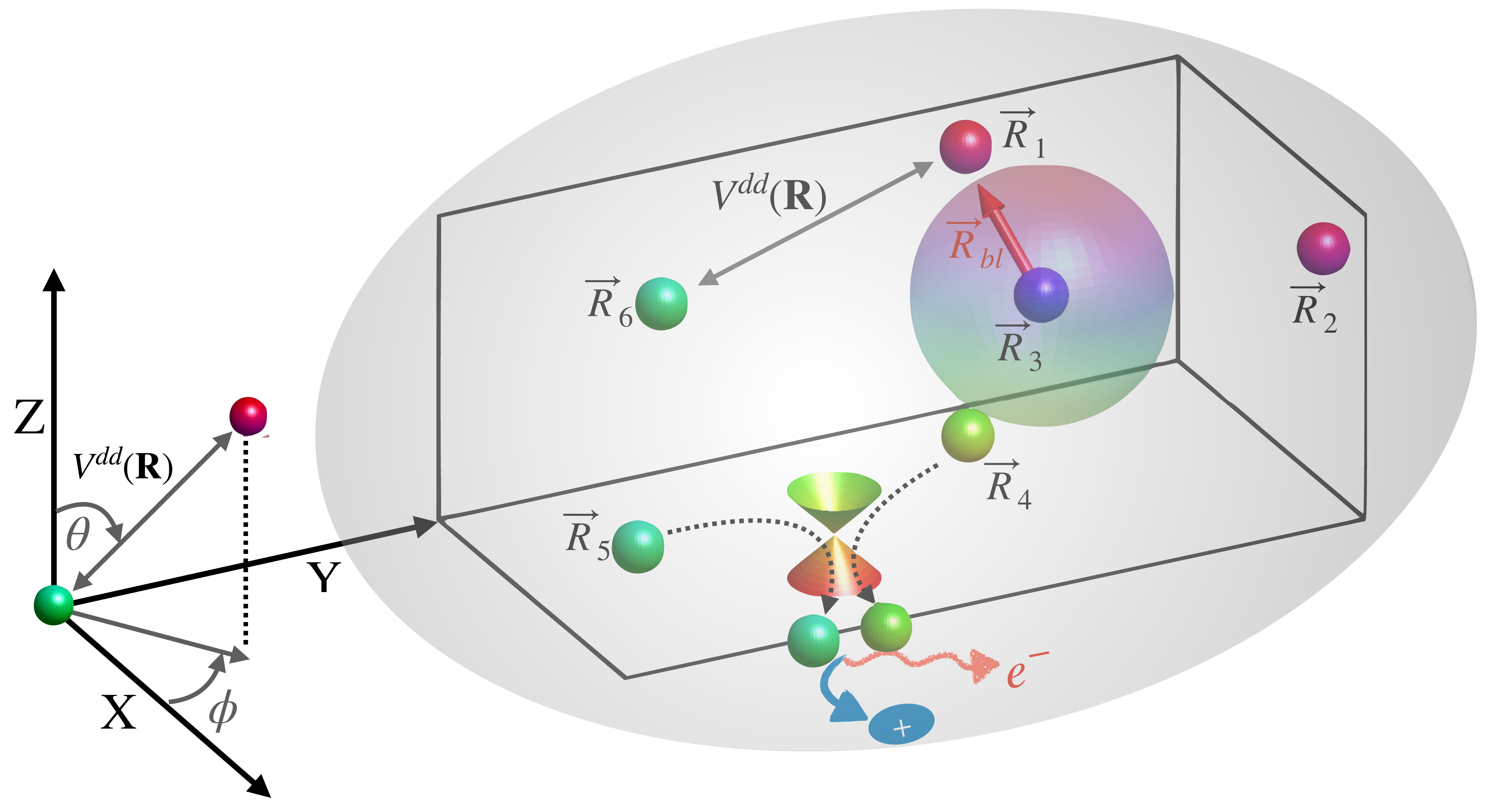}
\caption{\label{sketch} Sketch of a randomly distributed assembly of Rydberg atoms (spherical balls) in a 3D bulk gas (gray shade) at locations $\mathbf{R}_1, ...., \mathbf{R}_6$. $R_{bl}$ is the blockade radius shown by the smaller transparent sphere, representing the initial minimal distance. After passing through a conical interaction (double cones) or region of strong non-adiabatic coupling, 
atomic repulsion can be overcome and the collision of two Rydberg atoms (here $4$ and $5$) results in ionization.}
\end{figure}
More complex dynamics arises when transition dipole-dipole interactions involve a large number of atoms, giving rise to collective exciton states with a delocalized excitation \cite{cenap:motion,wuster2018rydberg}. While a single Rydberg dimer possesses only strongly repulsive or attractive electronic states, the multi-atom system has states in which the resonant DD contribution vanishes for the closest pair of atoms. 
If the remaining repulsive van-der-Waals interactions are too weak to counteract kinetic energy initially gained from dipole-dipole repulsion, atoms paired  in these states can  collisionally ionize if they hit a neighboring Rydberg atom.

An additional feature of collective dipole-dipole interactions are conical intersections (CIs) between adjacent electronic states of the multi-atom Rydberg assembly \cite{wuester:CI}. Near a conical intersection, non-adiabatic transition necessarily become likely. Even if initialised in a state in which nearest atoms should repel, non-adiabatic transitions or passages through conical intersections can  transfer their electronic state to a non-repelling one. After that transfer, Rydberg atoms can ionize. We demonstrate that, for a suitable choice of parameters, most ionization events can be traced back to at least one non-adiabatic transition implying conversely that  ionization is an easily accessible experimental signature for a non-adiabatic transition. In comparison to earlier proposals to investigate conical intersections in Rydberg systems \cite{wuester:CI,leonhardt2014switching, leonhardt2016orthogonal, leonhardt2017exciton}, the scenario discussed in this article lifts the challenges of constraining the motion of atoms through trapping, tightly localizing the Rydberg excitation and high resolution Rydberg position measurements. 

Nonadiabatic effects due to conical intersections play a key role in many quantum chemical processes \cite{dantus2004introduction} such as photochemistry of vision \cite{hahn_rhodopsin} or DNA protection from UV radiation damage \cite{Perun_DNA_radiationless_JACS}. While computational methods can nowadays provide an impressive level of detail for  multi-dimensional nuclear wavefunctions of molecules \cite{burghardt:MCTDH_pyrazine,Worth-Burghardt-UsingMCTDHwavepacket-2008}, such detail may be hard to achieve  when interrogating molecules experimentally. More detail may be accessible, when studying dynamics around conical intersections  with cold atoms or molecules \cite{wallis2009conical,moiseyev:CI_lattice,zhu:diracfermions_lattice,larson:jahntellerspinorbit}, Rydberg atoms \cite{wuester:CI,wuster2018rydberg} or ions \cite{weibin:ion:CI}.

This article is organized as follows: In \sref{model}, we introduce the model for collective dipole-dipole interactions of Rydberg atoms, 
 the quantum-classical propagation scheme employed and our phenomenological model of the excitation sequence. In \sref{Position_dynamics} we  explore non-adiabatic dynamics with individual trajectories of the classical motion of Rydberg atoms, to clearly connect ionisation with non-adiabatic transitions. Finally, in \sref{Exciton_spectrum_most_probable}, we portray the dynamics in the trajectory averaged energy spectrum of the system.

\section{Dipole-dipole interactions in 3D Rydberg Gases} 
\label{model}

We consider an assembly of $N$ Lithium Rydberg atoms each with mass $M\approx 11000$ au.~and principal quantum number $\nu$. All atoms can move freely in 3D, with their collective positions described by the $3N$-component vector $\mathbf{R} = (\mathbf{R}_1,\cdots,\mathbf{R}_N)^T$, where $\mathbf{R}_k$ is the position of atom number $k$. We are interested in the joint effect of DD and vdW interactions.  Hence, besides constraining $\nu$, we consider angular momentum states $\ket{s(l=0)}$ and $\ket{p(l=1),m}$. For scenarios with only a single Rydberg atom  in a $\ket{p}$ state, the collective electronic basis can then be written as  $\{ \ket{\pi_n(m)} \}$,  where,
\begin{eqnarray}\label{Diabatic_electronic_basis}
\ket{\pi_n(m)} = \ket{s...(p,m)...s},
\end{eqnarray}
represents the state where the $n$'th atom carries the p-excitation with magnetic quantum number $m$, and all other atoms are in the $\ket{s}$ state.
We thus neglect spin-orbit coupling, a reasonable approximation 
for light atoms such as Lithium.

The resulting Hamiltonian for our system reads
\begin{eqnarray}\label{Electronic_Hamiltonian}
\hat{H} = - \frac{\hbar^2}{2M} \nabla_{\mathbf{R}}^2 + \hat{H}_{el}(\mathbf{R}),
\end{eqnarray}
where the first term accounts for the kinetic energy of the Rydberg atoms and  $\hat{H}_{el}(\mathbf{R}) = \sub{\hat{H}}{dd}(\mathbf{R}) + \sub{\hat{H}}{vdw} (\mathbf{R})$, with the resonant dipole-dipole interactions 
\begin{align}\label{Dipole_interaction}
\hat{H}_{dd}(\mathbf{R}) = \sum_{\stackrel{n, n^{\prime}=1}{m,m^{\prime}} }^{N}  V^{(dd)}_{m m^{\prime}}(\mathbf{R}_{n, n^{\prime}})
 \ket{\pi_{n}(m)} \bra{\pi_{n^{\prime}}(m^{\prime}) }\,.
\end{align}
Here, $V^{(dd)}_{m m^{\prime}}(\mathbf{R}_{n, n^{\prime}})$ are the matrix elements for the dipole-dipole interactions between atoms $n$ and $n^{\prime}$.
 They depend on $\mathbf{R}_{n n^{\prime}}{=}\mathbf{R}_n {-} \mathbf{R}_{n^{\prime}}$ and are explicitly given by \cite{robicheaux2004simulation}
\begin{align}\label{Dipole_interaction_matrix}
V^{(dd)}_{m m^{\prime}}(\mathbf{R}_{n n^{\prime}})&=-\sqrt{\frac{8\pi}{3}} \frac{\mu^2}{4\pi\epsilon_0 {R}_{nn^{\prime}}^3} (-1)^{m^{\prime}} \\
&\begin{pmatrix}
1 & 1 & 2 \\
m & -m^{\prime} & m^{\prime}-m
\end{pmatrix}
{Y}_{2,m^{\prime}-m} (\theta_{n n^{\prime}}, \phi_{n n^{\prime}}),
\nonumber
\end{align}
where $R_{n n^{\prime} } = |\mathbf{R}_{n n^{\prime}}|$, $\theta_{n n^{\prime}}$ and $\phi_{n n^{\prime}}$ are the polar angle and azimuthal angle of the inter atomic distance vector, respectively, as shown in \fref{sketch}, and the six numbers in the parentheses denote the Wigner $3-j$ symbol.  Note that resonant dipole-dipole interactions are in general anisotropic 
due to the dependence on $\theta$ and $\phi$ in \bref{Dipole_interaction_matrix} and mix the different azimuthal sublevels $m$. The anisotropy and the azimuthal state mixing could be suppressed using Zeeman shifts \cite{leonhardt2016orthogonal}, however, here we are interested in the pristine scenario.

The term  $\sub{\hat{H}}{vdw} (\mathbf{R})$ in $\hat{H}_{el}(\mathbf{R})$ pertains to the non-resonant van-der-Waals interaction
\begin{eqnarray}\label{vdW_interaction}
\hat{H}_{vdw}(\mathbf{R}) = - \frac{1}{2} \sum_{n, n^{\prime}=1}^{N} \frac{C_6}{R_{n n^{\prime}}^6} \mathbb{I},
\end{eqnarray}
where $\mathbb{I}$ is the unit operator in the electronic space and $C_6$ is the dispersion coefficient which characterizes the strength of van-der-Waals interactions 
dependent on the principal quantum number $\nu$. Throughout this article, we shall assume atoms in $
\nu=80$ and hence $C_6 = -1.7 \times 10^{22}$ au, from \cite{singer2005long} and $\mu=4096$ au~in \bref{Dipole_interaction_matrix}.

A full fledged quantum mechanical simulation of the time dependent Schr\"odinger equation for the Hamiltonian defined in \eref{Electronic_Hamiltonian} is computationally not possible for useful atom numbers $N$. Hence we employ Tully's surface hopping algorithm \cite{tully1990molecular, mobius2011adiabatic}, a quantum-classical method in which the motion of the Rydberg atoms is simulated classically using Newton's equation of motion,
\begin{eqnarray}\label{Newtons_EoM}
M \frac{\partial^2 \mathbf{R}}{\partial t^2} = - \nabla_{\mathbf{R}} U_{s} (\mathbf{R}(t)),
\end{eqnarray}
where $s$ is the index of the Born-Oppenheimer (BO) surface $U_{k=s}(\mathbf{R}(t))$ on which the Rydberg system is presently evolving. The surfaces follow from the eigenvalue equation
\begin{eqnarray}\label{Adiabatic_eval_equation}
\hat{H}_{el} (\mathbf{R}) \ket{\varphi_k (\mathbf{R})} = U_{k}(\mathbf{R}) \ket{\varphi_k (\mathbf{R})},
\end{eqnarray}
where we order eigenstates according to increasing  energy with increasing index $k$.

In contrast to the positions, the electronic state of the Rydberg assembly $ \ket{\psi(t)} = \sum_{nm} {c}_{nm}(t)  \ket{\pi_n(m)}$ is evolved quantum mechanically
with Schr\"odinger's equation
\begin{eqnarray}\label{Diabatic_SE}
i \hbar \frac{\partial}{\partial t} {c}_{nm}(t) =  \sum_{n'm'}\bra{\pi_n(m)} \hat{H}_{el} (\mathbf{R}) 
\ket{\pi_{n'}(m')} {c}_{n'm'}(t).
\end{eqnarray}
It is instructive to also express \bref{Diabatic_SE} in the adiabatic basis $\ket{\psi(t)} = \sum_{k} \tilde{c}_k(t) \ket{\varphi_k{(\mathbf{R} )}}$ formed by solutions of \bref{Adiabatic_eval_equation}. In terms of this basis the TDSE reads 
\begin{eqnarray}\label{Adiabatic_SE}
i \hbar \frac{\partial}{\partial t} \tilde{c}_k(t) = U_k(\textbf{R}(t) ) \tilde{c}_k(t) - i \hbar \sum_{l} d_{kl}(t) \tilde{c}_l(t),
\end{eqnarray}
with non-adiabatic coupling vectors
\begin{eqnarray}
\label{d_mn}
d_{kl} \approx - \frac{1}{M} \bra{\varphi_k(\textbf{R})} \pmb{\nabla}_{\mathbf{R}} \ket{\varphi_l(\textbf{R})} \cdot \frac{\partial \mathbf{R}}{\partial t}.
\end{eqnarray}
The latter couple the different BO surfaces, which is implemented into the motional dynamics \bref{Newtons_EoM} of 
 Tully's method by allowing stochastic jumps of the index $s\rightarrow s'$ with a probability set by $d_{ss'}$. The adiabatic and the diabatic coefficients are connected by the relation $\tilde{c}_k = \sum_{nm} c_{nm} \braket{\varphi_k{(\mathbf{R})} } {\pi_n(m)}$. Our simulations employ Tully's method with \eref{Diabatic_SE} coupled to \eref{Newtons_EoM}, but one can refer to \eref{Adiabatic_SE} for understanding non-adiabatic transitions.

\subsection{Excitation process and blockade}
\label{Initial_state}

The initial excitation of Rydberg atoms to states involving $\ket{\pi_n(m)}$ is a two step process. Firstly, within a bulk ultra-cold gas, ground state atoms are excited to the Rydberg state $\ket{s}$, typically using a two-photon excitation \cite{loew:rydguide:jpbreview}. Due to the strong vdW interactions between Rydberg atoms in the state $\ket{s}$, this step is affected by the dipole-blockade, which prohibits the excitation of more than one atom within a sphere with blockade radius $R_{bl}$. We estimate $R_{bl}$ as the distance at which the strength of van der Waals interactions becomes equal to the broadened linewidth of a laser with Rabi frequency $\sub{\Omega}{las}$ \cite{PhysRevLett.87.037901, urban2009observation}, resulting in
\begin{eqnarray}\label{Blockade_radius}
R_{bl} = \left( \frac{|C_6|}{ \sub{\Omega}{las}} \right)^{1/6}\,.
\end{eqnarray}
 For our simulation we assume $\sub{\Omega}{las}/(2\pi)\approx 60$ MHz, which for $\nu=80$ results in $R_{bl} = 5.9$ $\mu m$. The bulk excitation and blockade are then phenomenologically taken into account by drawing random positions from a sphere of radius $R=4 R_{bl} $  in 3D, and discarding those with $R_{nm}<R_{bl} $ for any pair of Rydberg atoms. A more sophisticated approach could model the excitation process using classical rate equations \cite{cenap_rateequation_PRA}. 
 After completing this first step of the process, we have an assembly of $N$ Rydberg atoms all in the state $\ket{s}$ (hence in a many-body state $\ket{s..s..}$) at random positions $\sub{\mathbf{R}}{ini}$ consistent with the excitation blockade.
 
In the second step we induce  a $p$-excitation to enable resonant dipole-dipole interactions with a
near resonant microwave pulse to $\ket{\pi_n(m)}$, linearly polarized along the quantisation axis. The corresponding Hamiltonian reads
\begin{align}
\label{Microwave_Hamiltonian}
\hat{H}_{mw}& = \sum_{n} \frac{\Omega }{2} \Big(\ket{\pi_n(0)} \bra{s} + \mbox{c.c.} \Big)\CR
& -\sum_{n}   \Delta\ket{\pi_n(0)}\bra{\pi_n(0)},
\end{align}
where $\Omega$ is the Rabi-frequency of the microwave and $\Delta$ its detuning from the bare $sp$ transition.
By this detuning, it is possible to selectively excite a chosen exciton state $\ket{\varphi_k}$ in \bref{Adiabatic_eval_equation}, while avoiding more than one $p$-excitation \cite{moebius:cradle}.

In practice, after atoms have been excited to Rydberg $s$-states at random initial positions $\sub{\mathbf{R}}{ini}$ as discussed above, the relative excitation probability $P_r(k)$ of state $\ket{\varphi_k(\sub{\mathbf{R}}{ini})}$ will depend on two factors: (i) The transition matrix element between $\ket{s..s..}$ and $\ket{\varphi_k(\sub{\mathbf{R}}{ini})}$, which depends on the microwave polarisation direction and (ii) the microwave frequency (detuning). We incorporate both effects phenomenologically in Tully's algorithm by randomly starting the simulation in exciton state $k$ with relative probability 
\begin{eqnarray}
\label{Rel_trans_probability}
P_r(k) ={\cal N} e^{-\frac{(U_k - \Delta)^2}{2 \sigma_U^2}} \tilde{P}_r (k)\,,
\end{eqnarray}
where $U_k$ are the exciton energies defined in \eref{Adiabatic_eval_equation}, $\Delta$ is the microwave detuning, $\sigma_U$ is the microwave linewidth and $ \tilde{P}_r (k)$ is a polarisation dependent factor, discussed in  \aref{Microwave_appendix}, that takes into account the matrix element. The normalisation factor ${\cal N}$ ensures $\sum_k P_r(k) =1$, and trajectories with $|U_k - \Delta|>2 \sigma_U$ for all $k$ are discarded. 
 Importantly, $P_r(k) $ contains a twofold dependence on the initial positions of Rydberg atoms after they have been excited to $\ket{s..s..}$: Through the exciton energies $U_k$ and the interplay of microwave polarisation and locations encoded in $\tilde{P}_r (k)$.

\section{Nonadiabatic dynamics from random initial positions}
\label{Position_dynamics}
\begin{figure}[htb]
\includegraphics[width=0.99\columnwidth]{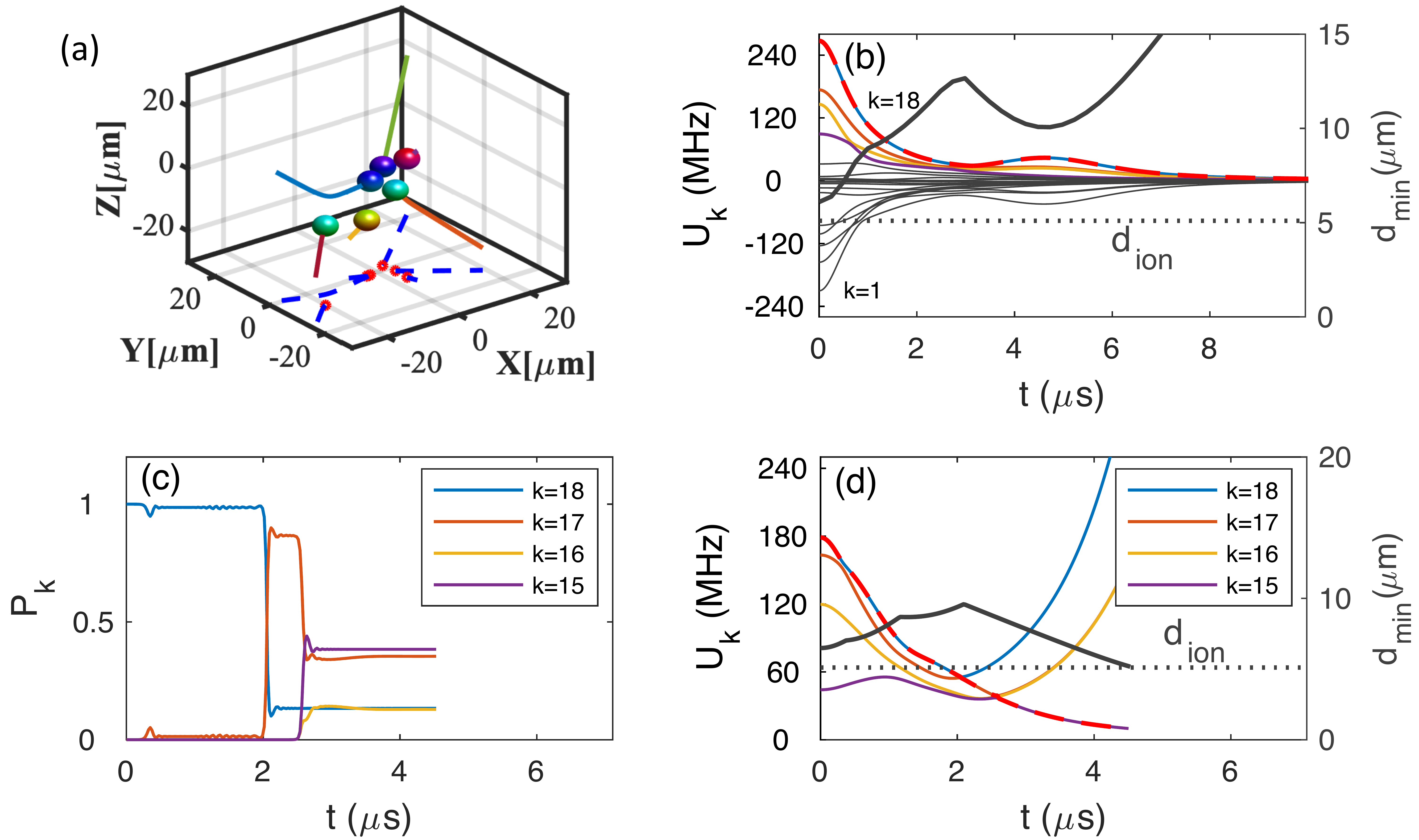}
\caption{\label{CIcrossing_k18} Electronic and motional Rydberg dynamics for the system prepared in the highest energy electronic state. (a,b) Trajectory without non-adiabatic transition. (c,d) Trajectory with non-adiabatic transition. (a) Initial configuration (colored balls) and trajectories (solid lines) of the atomic positions $\mathbf{R}_k(t)$. Red circles and blue dashed lines show the projection of initial positions and trajectories onto the XY plane. (b) Time evolution of potential energies $U_k(\mathbf{R}(t))$ and the minimum distance $\sub{d}{min}$ between atoms (thick gray line, right axis) without non-adiabatic coupling. $\sub{d}{ion}$ is the ionization distance defined in \eref{d_ion} (gray dotted line). (c) Adiabatic populations on the four highest energy repulsive surfaces, showing non-adiabatic dynamics. (d) Time evolution of potential energies $U_k(\mathbf{R}(t))$ similar to (b). The presently propagated surface $s$ in \eref{Newtons_EoM} is shown as red-dashed line, and the three highest potential energies and $\sub{d}{ion}$ are color coded as in (c). 
}
\end{figure}

%
We are now in a position to study the motional dynamics of $N=6$ Rydberg atoms, starting from initial locations $\sub{\mathbf{R}}{ini}$ as discussed in \sref{Initial_state} to explore how motion depends on the choice of the initially excited repulsive BO surface $k$ defined in \eref{Adiabatic_eval_equation}. Such a configuration is illustrated in \fref{CIcrossing_k18}(a), where the colored spheres represent the initial positions of Rydberg atoms. A single realization of the time evolution of atomic positions when the system is prepared in the highest energy repulsive state (i.e.~$k=3N=18$) is shown by the solid lines in 3D with the projection of trajectories onto the XY plane indicated by the dotted lines. \fref{CIcrossing_k18}(b) shows the time evolution of all the electronic energy surfaces $U_k(\mathbf{R}(t))$ on the left y-axis, together with the minimum distance $\sub{d}{min}=\mbox{min}_{nm}R_{nm}$ between the atoms, shown as thick grey line using the right y-axis. We have selected a trajectory without non-adiabatic transition, staying on the highest energy repulsive surface. Even when the initially accelerated atoms now encounter new collision partners, which happens around 
$t\approx5$ $\mu$s and causes  $\sub{d}{min}$ to exhibit a local minimum, the repulsion prevents a close encounter with ionisation.

A qualitatively different trajectory is shown in panel (c,d), where the system undergoes a sequence of non-adiabatic transitions from the highest energy repulsive surface to the surface with index $k=15$ as shown by the adiabatic populations in \fref{CIcrossing_k18}(c) and the red dashed lines in \fref{CIcrossing_k18} (d). The transitions are due to significant non-adiabatic coupling terms $d_{kl}$ in \eref{Adiabatic_SE}, usually when adjacent BO surfaces approach each other closely in energy. We find that on the surface $k=15$, atoms no longer repel, hence the the minimal distance $\sub{d}{min}$ decreases continuously after $\approx 2\mu$s when two Rydberg atoms encounter one another. If that happens, they would typically ionise, which is  phenomenologically modelled by declaring atoms collisionally ionized when they come closer than an ionisation distance $d_{\rm ion}$ where  the simulation is aborted, as shown in panels (c,d). We take $d_{\rm ion}$ as the distance below which the Rydberg energy spectra become dense when taking all electronic states into account. For such close distances, our effective state model based solely on $\ket{s}$ and $\ket{p}$ would break down. A rough estimate of $d_{\rm ion}$ is provided by the formula \cite{wuster2018rydberg}
\begin{eqnarray}\label{d_ion}
\sub{d}{ion}(n) = 2 \left(\frac{\mu^2 n^3}{\Delta E_{pd}(n)}\right)^{1/3},
\end{eqnarray}
where $\Delta E_{pd}$ is the energy difference between $\ket{p}$ and the nearest Rydberg state not included in the model, $\ket{d(l=2)}$. For the example shown in \fref{CIcrossing_k18}, the non-adiabatic transition therefore ultimately leads to ionization of the Rydberg atoms after $t=4$ $\mu s$, much earlier than the lifetime $\tau\approx 50$ $\mu$s of our $6$-atom system \cite{footnote:lifetime, wuster2018rydberg}. 

\begin{figure}[htb]
\includegraphics[width=0.99\columnwidth]{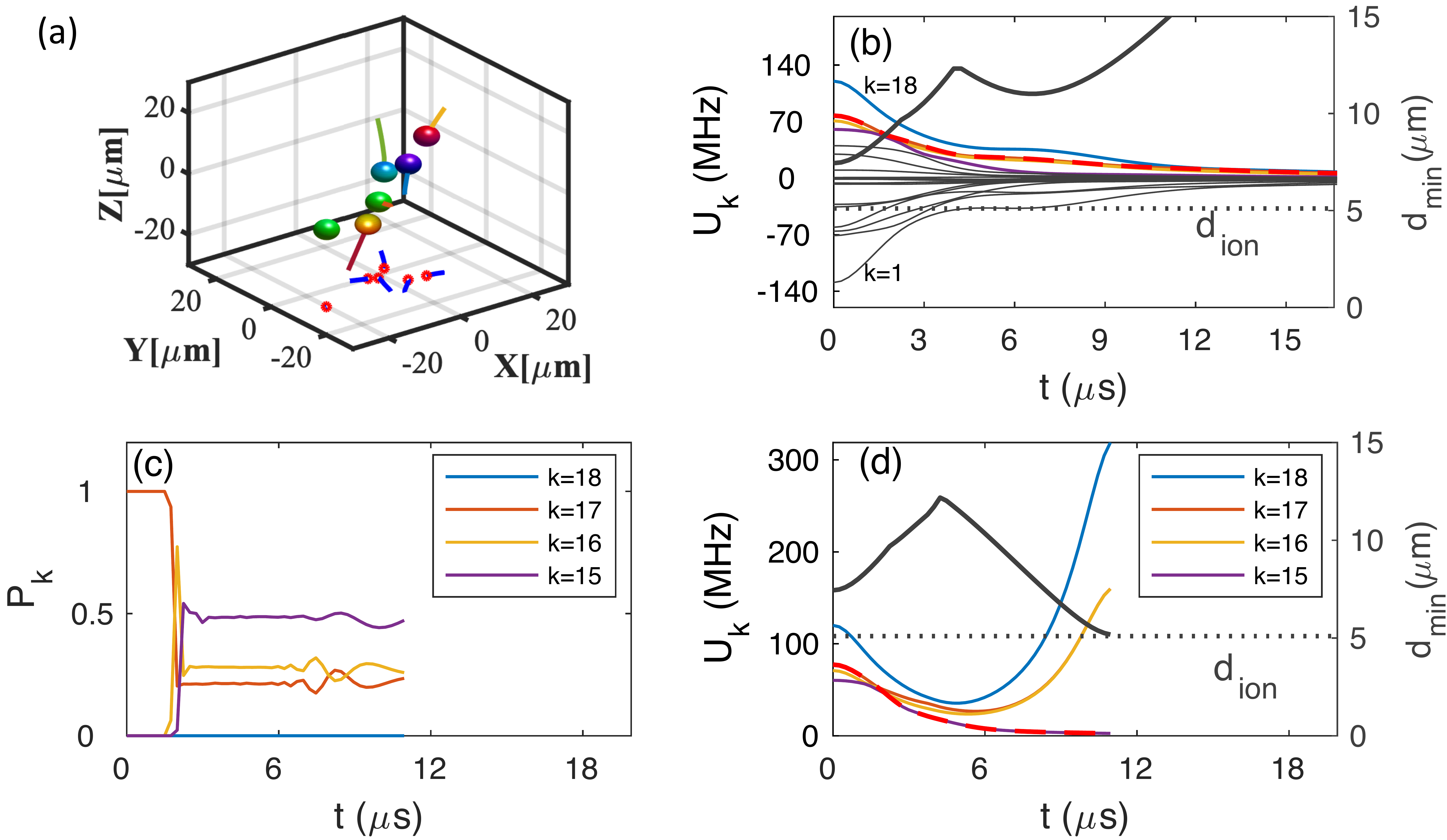}
\caption{\label{CIcrossing_k17} The same as \fref{CIcrossing_k18}, but starting from the
the second most energetic electronic eigenstate $k=17$, instead of the highest energy one, $k=18$.}
\end{figure}

Next we show, that qualitatively similar dynamics can be found when the Rydberg assembly is initialised on the next few less energetic energy surfaces, starting in 
 \fref{CIcrossing_k17}  from $k=17$. The figure illustrates the atomic dynamics up to the ionisation time, in (a) together with the time evolution of the electronic energy eigenstates and in (b,d) with the minimum distance between atoms. We see in \fref{CIcrossing_k17} (c,d) that due to the non-adiabatic coupling the system can quickly jump from surface $k=17$ to $k=15$, where it is later ionized. To underscore the need for non-adiabatic transitions, we also show in \fref{CIcrossing_k17} (b) the time evolution of eigenstates and minimal distance when non-adiabatic jumps of the surface index $s$ in \eref{Newtons_EoM} are disabled, but using the same initial configuration as in (c,d). As a consequence, the system remains on the surface $k=17$, maintaining  repulsive dynamics without ionization.

The scenario is qualitatively similar when starting on the surface $k=16$. In all cases, the first surface for which repulsion is lost and the Rydberg atoms can ionize is $k=15$ for $N=6$.  Consequently, we find that when starting on $k=15$, ionisation is possible even without a prior non-adiabatic transition. This is not the case from any of the higher surfaces.

Our inspection of single trajectories starting on the three highest energy surfaces with indices  $k=16,17,18$ has revealed that  ionization of Rydberg atoms must be preceded by at least one non-adiabatic transition. Therefore,
the observation of  collisional ionization of Rydberg atoms after initialising the system repulsively can be used as a signature of non-adiabatic transitions in the experiment.

\section{Averaged  dynamics}
\label{Exciton_spectrum_most_probable}
The single trajectory simulations in \sref{Position_dynamics} from selected electronic states provide a more detailed picture of the dynamics, but they would not be individually experimentally accessible. Instead one has to average over many repetitions of an experiment and address exciton states through the microwave detuning, corresponding to multi-trajectory averages with random initial electronic states according to \bref{Rel_trans_probability}. We present these averaged simulations in this section. 
 
\begin{figure}[htb]
\includegraphics[width=0.99\columnwidth]{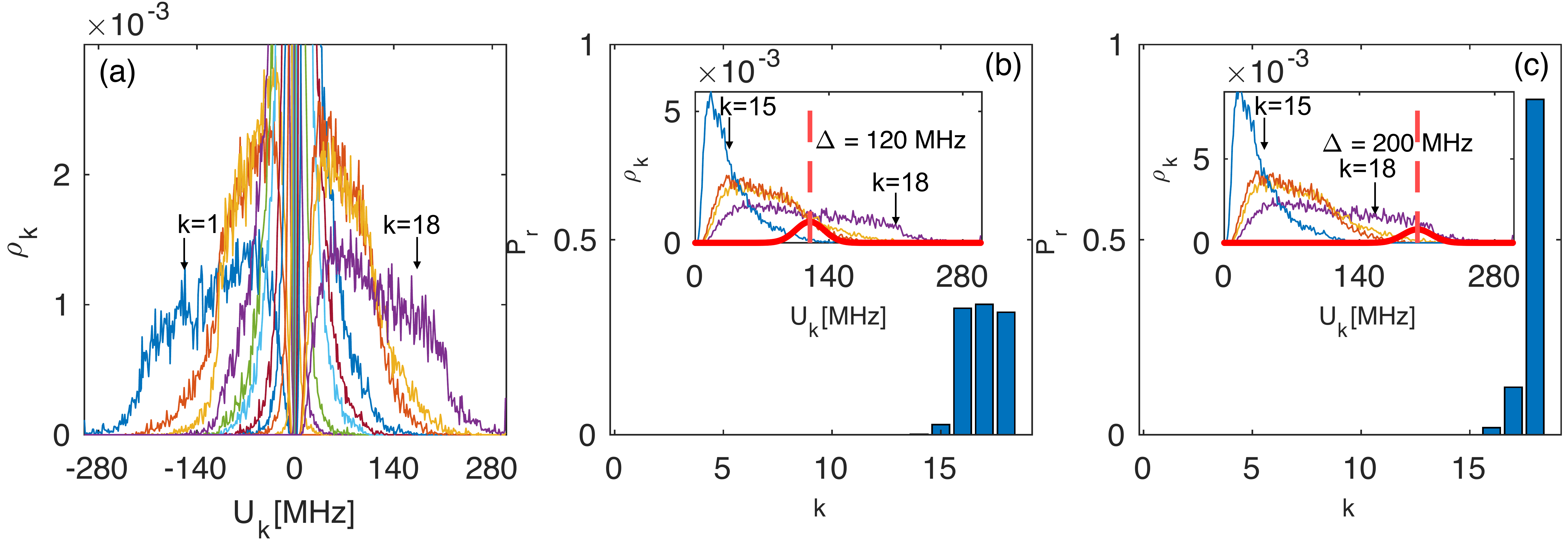}
\caption{\label{Energy_distribution_combine}(a) Probability distribution $\rho_k$ of energies $U_k$, $1\leq k\leq 18$, for 5000 random configurations ($\mathbf{R})$ of atoms. We show a zoom on the high energy flank in (b,c), together with a certain microwave line profile $P_r$ (thick red line) with detuning (b) $\Delta=120$ MHz and (c) $\Delta=200$ MHz respectively. The inset shows the resultant surface index distribution after excitation, according to \eref{Rel_trans_probability}.
}
\end{figure}

 In the proposed experiment with random Rydberg excitation location in a bulk gas, initialising a specific exciton state $k$, will pose practical challenges. The most straightforward approach would be detuning a microwave as discussed in \sref{Initial_state} such that it is resonant with that state: $\Delta=U_k$. However, the exciton energies $U_k$ depend on  all Rydberg positions $\sub{\mathbf{R}}{ini}$, which are random subject to constraints by the blockade. As an initial step to address this challenge, we show the histogram of initial exciton energies for those random positions in \fref{Energy_distribution_combine}.
 While the energy distribution of energetically neighboring excitons usually overlap,  we can see that the tails contain energy regions where only the highest energy surface is present. 
 Choosing a detuning in that region, e.g.~$\Delta = 200$ MHz and narrow line-width  $\sigma_U = 15.0$ MHz, we can  achieve excitation of the Rydberg assembly almost entirely on the highest energy surface, as shown in \frefp{Energy_distribution_combine}{c}. For realisations of positions in which the atoms are too far apart to provide an exciton with these high energies, no excitation would happen for this detuning in an experiment. We model only the cases where excitation of a $\ket{p}$ state is successful. For the distribution of excited surfaces we phenomenologically model the microwave excitation probability \eqref{Rel_trans_probability}  with more details in \aref{Microwave_appendix}.
 Lowering the detuning to $\Delta = 120$ MHz and $\sigma_U = 15.0$ MHz allows a tuning of the distribution of surfaces $k$, with almost equal contributions of the three highest ones, $k=16,17,18$.
 
\begin{figure}[htb]
\includegraphics[width=0.99\columnwidth]{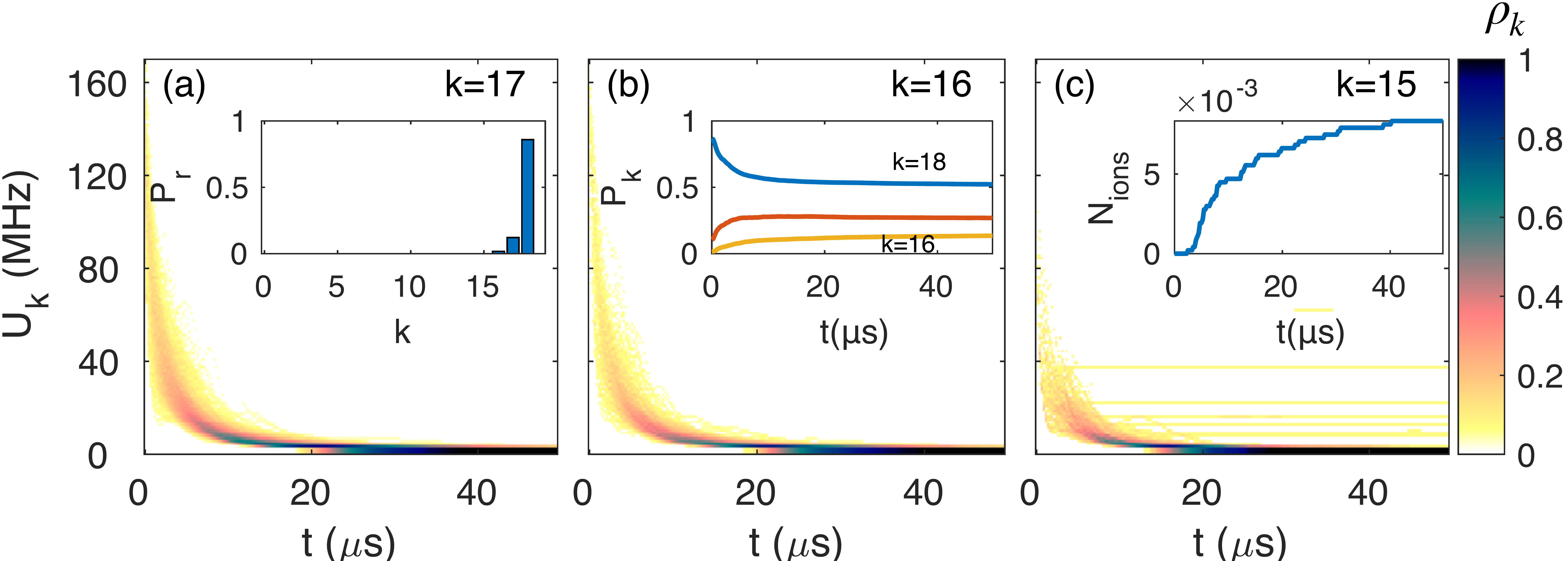}
\caption{\label{Energy_density_D200_v1}  Mean time-resolved potential energy density on the (a) second, (b) third and (c) fourth most energetic BO surface, starting with the microwave detuned to $\Delta = 200$ MHz so that the system most likely begins on the first surface. To emphasize low density features, we plot the square root of energy densities and adjust the colorbar range. The insets show (a) the relative initial excitation probability of each surface, (b) adiabatic populations and (c) the mean number of ions per trajectory as a function of time.
}
\end{figure}

We now present multi-trajectory surface-hopping simulations starting from random initial state distributions as shown in \fref{Energy_distribution_combine}, using $\Delta = 200$ MHz such that the probability of excitation onto the highest energy repulsive surface is $90 \%$, see  inset of \fref{Energy_density_D200_v1}(a). The figure shows the time and surface resolved potential energy density $\rho_s(t)$, which we construct by binning the potential energy $U_s(t)$ of the currently propagated BO surface $s(t)$ regarding energy and surface, and by averaging the result over all the trajectories, see also \cite{leonhardt2017exciton}.

\begin{figure}[htb]
\includegraphics[width=0.99\columnwidth]{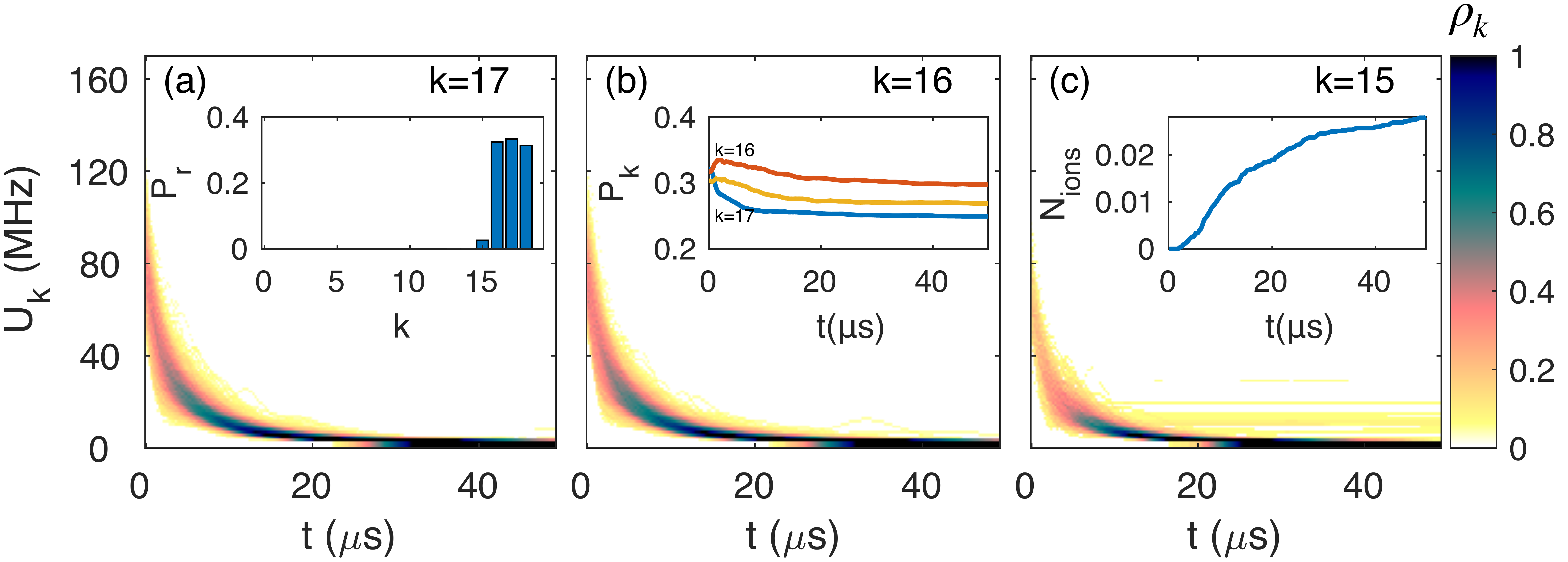}
\caption{\label{Energy_density_D120_v1}
Mean time-resolved potential energy density similar to \fref{Energy_density_D200_v1} but for the microwave detuning of $\Delta = 120$ MHz, such that more surfaces are populated initially.
}
\end{figure}

We see, that from the highest energy surface, the system jumps non-adiabatically to lower energy surfaces and then ultimately can be ionized after reaching the surface with index $k=15$. Since ionisation is implemented by stopping the time-evolution and sampling constant quantities thereafter, it shows up in these histograms as (unphysical) horizontal stripes, which are to be taken simply as a pointer towards the ionisation event (where the stripes intersect the bulk distribution).
If the detuning is reduced to $\Delta = 120$ MHz with linewidth $\sigma_U = 15.0$ MHz, one significantly excites three of the highest energy surfaces as shown in \fref{Energy_distribution_combine}(b). This relative transition probability is again shown in the inset of \fref{Energy_density_D120_v1}(a). Starting from such an initial state, \fref{Energy_density_D120_v1} shows the evolution of time-resolved potential energy density for the three most strongly participating surfaces, similarly to \fref{Energy_density_D200_v1}. Since fewer non-adiabatic transitions are required on average to reach the ionising surface, this scenario exhibits an about three times higher ionisation signal than the one of  \fref{Energy_density_D200_v1}. Spectra as shown in \fref{Energy_density_D200_v1} and \fref{Energy_density_D120_v1} are experimentally accessible through microwave spectroscopy as in \cite{celistrino_teixeira:microwavespec_motion}.

Overall we have shown that for both these microwave detunings, ionization of Rydberg atoms is strongly linked to non-adiabatic transitions, since the system will be initialised solely on surfaces on which ionisation does not occur directly, as shown in \sref{Position_dynamics}. We also demonstrated that on the high energy tails of the initial energy distribution, a reasonable control over the initial exciton state is possible through the microwave detuning. The control over the initial state  translates into control over the ionization probability.

The rate of ionisation for the simulations presented here remains relatively low is an artefact of our restriction to just $N=6$ atoms to keep simulations tractable. 
For only six atoms it is relatively unlikely that a pair of atoms initially repels, the system then undergoes non-adiabatic transitions to leave the repelling surfaces, and an atom subsequently still encounters a new collision partner on its outwards journey in order to ionize. The situation would be very different in an experiment, where a much larger number of Rydberg atoms can be easily excited. If the initially repelling atoms are surrounded in all direction by a larger number of Rydberg atoms, it is reasonable to expect that the ionisation probability after a non-adiabatic transition could approach unity. 

\section{Conclusions} 
\label{Conclusion}
%
We have modelled the joint electronic and motional dynamics of an assembly of few ($N=6$) Rydberg excitations that are created at random 3D positions in an ultra-cold gas. Atoms subsequently move according to resonant dipole-dipole interactions.
With a simple phenomenological model for microwave excitation with a fixed detuning, we have shown that experiments can selectively initialise this motional dynamics in dipole-dipole exciton eigenstates that have entirely repulsive character. We have modelled the motion of the Rydberg assembly with a quantum classical surface hopping algorithm, that permits non-adiabatic transitions between exciton states.

These simulations reveal frequent non-adiabatic transitions for the parameters selected, rendering it likely that the group of Rydberg atoms  reaches a Born-Oppenheimer surface that is no longer repulsive and therefore permits collisional ionisation of Rydberg atoms. This causal chain of events turns an ion count into an experimentally accessible flag for non-adiabatic transitions.

Earlier simulations of non-adiabatic dynamics near conical intersections in Rydberg systems assumed trapped Rydberg atoms \cite{wuester:CI,leonhardt2014switching, leonhardt2016orthogonal} or tightly localized excitation beams \cite{leonhardt2017exciton}, 
and require high resolution Rydberg position measurements for the observation of the results. While posssible with a dedicated apparatus \cite{Lampen_magicwavelength_PRA,thaicharoen2015measurement,thaicharoen:dipolar_imaging,guenter:EITexpt}, these requirements pose a challenge to most ultra cold Rydberg experiments. In contrast, the results presented here should be observable using routinely applied random Rydberg excitation in a thermal gas and interrogation via ion counting.

\acknowledgments
We thank Karsten Leonhardt for contributions to the codes used and the Max-Planck society for financial support under the MPG-IISER partner group program. RP is grateful to the Council of Scientific and Industrial Research (CSIR), India, for a Shyama Prasad Mukherjee (SPM) fellowship for pursuing the PhD (File No. SPM-07/1020(0304)/2019-EMR-I).
\appendix

\section{Microwave transitions}
\label{Microwave_appendix}
The exciton states defined in \bref{Adiabatic_eval_equation} can be expressed explicitly in the basis \bref{Diabatic_electronic_basis} as
\begin{eqnarray}\label{Exciton_state}
\ket{\varphi_k} = \sum_{m,n} f^{(k)}_{nm} \ket{ \pi_n(m) }\,,
\end{eqnarray}
where $ f_{nm}^{(k)} = \braket{ \varphi_k }{ \pi_n(m) } $ are the component amplitudes in the basis state $\ket{ \pi_n(m) }$ for the system eigenstate $\ket{\varphi_k}$.  With \eqref{Exciton_state} the transition probability $P_k$ from the $\ket{s..s..}$ state to the exciton state \eref{Exciton_state}, using \eref{Microwave_Hamiltonian},  follows the proportionality
\begin{eqnarray}\label{P_k_appendix}
P_k \propto |\bra{\varphi_k} \hat{H}_{mw} \ket{s..s..}|^2 \propto \Big[\sum_{n} f_{n0}^{(k)} \Big]^*  \Big[\sum_{n} f_{n0}^{(k)} \Big]\,.
\end{eqnarray}
We are not interested in absolute probabilities for the simulation, since we only want to model trajectories where \emph{some} exciton state has been excited, but \bref{P_k_appendix} is sufficient to infer the relative 
transition probability $\tilde{P}_r (k)$ onto the BO surface $k$ as
\begin{eqnarray}\label{P_k_rel_appendix}
\tilde{P}_r (k) = \frac{P_k}{\sum_k P_k}.
\end{eqnarray}
Using \eref{P_k_rel_appendix} the relative transition probability for different microwave detunings $\Delta$ is shown in \fref{Energy_distribution_combine}(b) and (c). 
For simulations in \fref{Energy_density_D200_v1} and \fref{Energy_density_D120_v1} we remove trajectories for which all exciton energies are far from the microwave resonance, $|U_k - \Delta|>2 \sigma_U$ $\forall k$, since when using an absolute excitation probability these cases would simply not show any excitation of Rydberg $p$ states.



\begin{thebibliography}{43}
\bibitem[{foo()}]{footnote:lifetime}
\bibinfo{note}{The lifetime of $N$ Rydberg atoms $\tau_{n,l}$ at zero
  temperature can be estimated by $ \tau_{n,l} = \tau_0^{(l)} n^3/N$, where
  $\tau_0^{(l)}$ is the lifetime of a single Rydberg atom. For $n=80$ we have
  $\tau_0^{(l)}\approx 0.058 $ $\mu s$ from \cite{beterov2009quasiclassical},
  hence for $N =6$, $\tau_0^{(l)}\approx 50 \mu s$}.
\end{thebibliography}
\end{document}